\documentclass[12pt]{article}

\usepackage{graphics}
\usepackage{amssymb}
\usepackage{graphics}
\usepackage{amssymb}
\usepackage[driver]{graphicx}

\newcommand{\ra}{\rightarrow}
\newcommand{\bra}{\langle} \newcommand{\ket}{\rangle}
\newcommand{\be}{\begin{equation}}
\newcommand{\ee}{\end{equation}}
\newcommand{\bea}{\begin{eqnarray}}
\newcommand{\eea}{\end{eqnarray}}
\newcommand{\eps}{\epsilon}

\newcommand{\ffi}{\varphi}
\newcommand{\QED}{\mbox{\rule[-1.5pt]{6pt}{10pt}}}

\newcommand{\pd}{\partial}
\newcommand{\D}{{\rm d}}
\newcommand{\grintl}{[\kern-.18em [}
\newcommand{\grintr}{]\kern-.18em ]}

\newtheorem{lem}{Lemma}[section]
\newtheorem{prop}{Proposition}[section]
\newtheorem{thm}{Theorem}[section]
\newtheorem{cor}{Corollary}[section]

\def\R{\mathbb{R}}

\def\un{\hbox{$\mit I$\kern-.77em$\mit I$}}
\def\0{\hbox{$\mit I$\kern-.70em$\mit O$}}
\def\r{I\kern-.277em R}

\def\I{\mathbb{I}}

\begin{document}

\title{Magnetic transport in a straight \\ parabolic channel}
\author{P.~Exner$^{a,b}$, A.~Joye$^{c}$, and H.~Kova\v{r}\'{\i}k$^{a,c,d}$,}
\date{}
\maketitle

\begin{flushleft}
 {\em a) Department of Theoretical Physics, Nuclear Physics Institute,
 Academy \phantom{a) }of Sciences, 25068 \v{R}e\v{z} near Prague
 \\ b) Doppler Institute, Czech Technical University, B\v{r}ehov\'{a}
 7, 11519 Prague, \phantom{a) }Czech Republic
 \\ c) Institut Fourier, Universit\'{e} de Grenoble 1, 38402 Saint-Martin
 d'Heres, \phantom{a) }France \\
 d) Faculty of Mathematics and Physics, Charles University,
 \phantom{a) }V~Hole\v{s}ovi\v{c}k\'ach~2, 18000 Prague
 \\ \quad{\em exner@ujf.cas.cz, joye@ujf-grenoble.fr,
 kovarik@ujf.cas.cz}}
\end{flushleft}

\begin{abstract}
\noindent We study a charged two-dimensional particle confined to
a straight parabolic-potential channel and exposed to a
homogeneous magnetic field under influence of a potential
perturbation $W$. If $W$ is bounded and periodic along the
channel, a perturbative argument yields the absolute 
continuity of the bottom of the spectrum. We show it can have 
any finite number of open gaps provided the confining potential 
is sufficiently strong. However, if $W$ depends on the periodic
variable only, we prove by Thomas argument that the whole spectrum 
is absolutely continuous, irrespectively of the size of the 
perturbation.
On the other hand, if $W$ is small and satisfies a weak localization 
condition in the the longitudinal direction, we prove by 
Mourre method that a part of the absolutely continuous spectrum persists.
\end{abstract}

\section{Introduction}

The problem of magnetic transport goes back to the early eighties of
the last century \cite{halp, mas}. Then it was found that the
transport can be achieved in a system with a homogeneous magnetic
field if boundaries are present. These so called edge currents
found numerous applications in solid-state physics. Recently it
has been shown that that such a type of transport exists even when
the boundary is replaced by a periodic array of point obstacles
\cite{ut, ejk}; in this case the propagation along the array is a
purely quantum effect.

On the other hand, it was also recognized that a suitable
translationaly symmetric variation of the magnetic field itself
can induce transport. A simple and transparent example of such a
variation is provided by a step of the magnetic field intensity.
As with the conventional edge states, the propagation here can be
understood also at the classical level, since the cyclotronic
radius at both sides of the step is different -- see
\cite[Sec.~6.5]{cycon}. Similarly the transport can exist in the
case when the magnetic field has the same asymptotics in both
directions perpendicular to the field variation \cite{iwa, mant, ek}.

It is naturally of both theoretical and practical interest to understand 
how such a magnetic transport is influenced by various perturbations.
Recently several studies treated the problem of edge-current
stability with respect to a sufficiently weak ``random''
perturbation (i.e., a deterministic bounded potential of an
arbitrary shape). The particle was at that supposed to be confined
in a semi-infinite region by either a smooth potential wall which
vanishes in one half-plane and rapidly increases in the other
\cite{macr}, or by a Dirichlet boundary \cite{biev, fro}. The
proofs were based on commutator methods. In \cite{macr} it
was shown, using a version of the virial theorem, that in certain
parts of the spectrum the Hamiltonian of the particle cannot have
any eigenstates, so that the spectrum is there purely continuous.
In \cite{biev, fro} the Mourre theory of positive commutators was
used to prove that for energy intervals away of the Landau levels
the spectrum remains purely absolutely continuous, i.e. that the
transport survives in the presence of an impurity potential.
Moreover, the argument of \cite{fro} works under weaker conditions
and extends the result to more general planar domains containing
an open wedge.

On the other hand, much less is known about the situation when the
particle is confined from both sides. It is true, of course, that
many numerical studies of such systems which model various quantum
wires can be found in the physical literature, but rigorous
results are scarce. This is our motivation to consider such a
potentially confined channel. For the sake of simplicity we
suppose that the channel is straight and the potential is
parabolic with constant strength along the axis. This is
certainly a reasonable model which has the advantage that it
allows us to solve the unperturbed problem analytically. We prove
two types of results.

First, if a bounded potential $W$ periodic in the longitudinal
direction is added, the bottom of the spectrum remains absolutely continuous
for weak enough perturbations. On one hand, we discuss the number
of gaps in such a continuous spectrum as a function of the
strength of the confining potential. On the other hand, we prove
that if $W$ depends only on the longitudinal or on the transverse
variable, the whole spectrum remains absolutely continuous, independently
of the strength of the potential. 

Second, if the perturbation
$W$ is no longer assumed to be periodic, we prove that a part of the spectrum
remains absolutely continuous provided $W$ is small in a suitable
sense and satisfies a weak ``localization'' condition.

Let us describe in more details the results and the contents of
the paper. The unperturbed Hamiltonian will be
\be
 H_0 = H_L(B)+\omega^2\, y^2\,,
\ee
where $H_L(B)= p_y^2+(p_x+By)^2$ is the free magnetic Hamiltonian
with homogeneous magnetic field $B$. The last operator
corresponds to the Landau gauge, which we will use throughout the
paper.

In the following two sections we analyze periodic
perturbations, i.e. the structure of the spectrum of
\be \label{fullHam}
 H = H_0+W \ee
where the potential $W(x,y)$ is $\ell$-periodic in $x$. The
periodicity enables us to use the Bloch decomposition and to write
the generalized eigenfunctions of $H_0$ in the form
\be
\psi_m(x)\ffi_n(y,m+\theta)
\ee
where $m\in\mathbb{Z}$, $n\in{\bf N}_0$, and $\theta$ is the
corresponding Bloch parameter running through the Brillouin zone
$[-\pi/\ell,\pi/\ell)$. In the absence of perturbation it is
straightforward to see that the spectrum is purely absolutely
continuous and includes all energies in the interval
$[\sqrt{B^2+\omega^2},\infty)$. Perturbation theory then shows that
for any $E>0$, the part of the spectrum inside the interval 
$[\sqrt{B^2+\omega^2}-\| W\|, E]$
is still purely absolutely continuous, provided $\| W\|$ 
is small enough.

Next using an appropriately modified
Thomas argument -- cf.~\cite{tom} and the generalization in
\cite[Sec.~XIII.16]{rees} -- we will prove in Theorem \ref{thomas}
that the whole spectrum of 
$H$ remains purely absolutely continuous if $W(x,y)\equiv W(x)$
depends on $x$ only and is essentialy bounded. The same is
true if $W(x,y)\equiv W(y)$ depends on the transverse variable
only and is essentially bounded.

Finally, we address the question about the number of open gaps in the
spectrum. One can find a partial answer using properties of the
function $W_0:=(\ffi_0,W(\cdot,y)\ffi_0)$ which represents the
projection of the potential onto the lowest transverse mode. If
the latter is non-constant, the one-dimensional Schr\"odinger
operator $K=-\pd_x^2+W_0(x)$ on $L^2(\R)$ has by
\cite[Thm.~XIII.90]{rees} a purely absolutely continuous spectrum
with open gaps -- at least one but generically infinitely many. We
will show in Section \ref{opengaps} that these gaps persist in the
spectrum on the operator (\ref{fullHam}) provided the coupling
constant of the confinement is large enough, see Theorem \ref{gaps}. 
Therefore, such a channel can have generically any finite number of 
open gaps for any bounded $x$-periodic perturbation, provided it is
confining enough.

Non-periodic perturbations require a different technique. In the
last part of this paper, Section \ref{scatt}, we address this
question in a similar way to that of the papers mentioned above,
namely by using a Mourre operator related to a distinguished
classical quantity. Recall that the central point of the Mourre
theory is to find a suitable self-adjoint conjugate operator $A$
such that in certain states the expectation value of $[H_0+W,iA]$
will have a definite sign. Classically, it amounts to finding an
observable increasing in time. This motivated the choice of the
conjugate operator in \cite{biev,fro} where the classical particle
followed the boundary counterclockwise and therefore propagated in
a definite direction. Accordingly, the coordinate parallel to the
boundary gave a conjugate operator with the needed properties.

By contrast, in our case there are two ``boundaries'' which allow
for classical motion in both directions along the $x$ axis. Of
course, they are edges with a grain of salt, since their
``distance'' depends on the particle energy.

Little is known so far about the stability of transport in
systems without a preferred direction. The existing results always
assume in some form that the ``opposite'' edge currents can be
placed at arbitrarily large distance to prevent their destructive
interference. This is the case for domains containing wedges in
\cite{fro} which we mentioned earlier. Another example is the
recent paper \cite{fem}, which studies the nature of the spectrum
of random Schr\" odinger operator with magnetic field in a finite
macroscopic system. The particle is supposed to be confined in one
direction by two smooth boundaries separated by a
distance equal to $L$, and the other direction is $L$ periodic. It
is then shown that for $L$ large enough there exist realizations
of random potentials such that the spectrum in the vicinity of
Landau levels contains both current carrying states and localized
states. Roughly speaking, this is due to decoupling of bulk and
edge states in the limit of large $L$. It is also announced, that away
from the Landau levels there are the current carrying states only. 
Notice that the transverse
distance in \cite{fem} may grow slower, say as $L^{\alpha}$ with
$\alpha\in (0,1)$, but it cannot be kept constant.

In models of a channel with a fixed cross section there is no
external parameter to control the decoupling, and it is not a
priori clear how the spectrum will behave. We start the Mourre
analysis by solving the classical problem in the absence of the
potential $W$. The trajectories turn out to be drifting ellipses.
We take the $x$-coordinate of the ellipse center multiplied by the
corresponding momentum component as the quantity to determine the
conjugate operator. This allows us to find that under suitable
smallness assumptions about $W$ there are intervals separated from
the modified Landau levels where the spectrum contains no eigenvalues
or even, under stronger hypothesis on $W$, 
remains absolutely continuous. Unfortunately, the assumptions include
finiteness of $\sup|x\, \pd_x W(x,y)|$ respectively $\sup|x^2\, \pd_x^2
W(x,y)|$ which can be regarded as a sort of localization
requirement. Of course, many ``non-local'' 
potentials fit in, say those with different limits as
$x\to\pm\infty$, and any powerlike decay at large $|x|$ will do,
however, the said condition excludes the most typical random
potentials in the form of a sum of randomly placed copies of a
single-impurity potential. For such potentials we establish the
existence of transport only in the situation when the ``dirty''
part of the channel has a finite length, see Theorem \ref{main}. We
also discuss the behaviour of our model in the limit of strong
confinement, i.e. when $\omega\ra\infty$. 

More than that, we show in Section \ref{gen} that any Mourre  
operator quadratic in the canonical variables will lead here to the
same restriction. Hence 
an attempt to establish for a ``fixed-width'' channel a result
comparable to \cite{biev,fro} by the conjugate-operator method has
to employ another $A$. Obvious candidates are those which combine
first order canonical variable with a (sign-changing)
localization of the particle in the vicinity of the edges.
However, attempts in this direction we are aware of have not been
successful so far and the problem remains open.

\section{Periodic perturbations} \label{period}

In this section we first give explicit expressions for the
eigenvalues and eigenfunctions of $H_0$, which is possible due to
the specific choice of our confinement potential. Then, as
mentioned above, we will investigate the nature of the spectrum
when we add a periodic perturbation.

The Hamiltonian of the system we are interested in is thus of the
following form,
\be\label{depart}
H =  -\pd_y^2+(-i\pd_x+yB)^2+\omega^2 y^2 + W(x,y) \quad {\rm on}
\quad \label{hamiltonian}  L^2(\R^2)\,, \ee
where $W$ is bounded and $\ell$-periodic in $x$. 
The scaling 
$$
x,y\ra \lambda\, x,\, \lambda\, y,\,
B\ra \lambda^{-2}\, B,\, \omega\ra \lambda^{-2}\,
\omega,\, W\ra \lambda^{-2}\, W
$$ 
gives  $H\ra \lambda^{-2}\, H$.  
Without loss we can thus assume $\ell=2\pi$.   
By
\cite[Thm.~X.34]{rees}, $H$ is  e.s.a. on $C_0^{\infty}(\R^2)$. We
use the periodicity of $W$ and apply the Bloch decomposition in
$x$ writing
\be
H = \int_{|\theta|\le 1/2}^{\oplus} H(\theta)\, \D\theta
\ee
where $H(\theta)$ has the form (\ref{hamiltonian}) on
$L^2([0,2\pi]\times\R)$
with the boundary conditions
\be
\partial_x^j\psi(2\pi -,y) = e^{i\theta 2\pi} \partial_x^j\psi(0+,y)\,,
\quad j=0,1\,. \label{bc} \ee
Let us now turn to the properties of the fiber operator
\be
\tilde{H}_0(\theta)= -\pd_y^2+(-i\pd_x+yB)^2+\omega^2 y^2\,. \ee
After transferring $\theta$ from the boundary conditions to the
operator we find that $\tilde{H}_0(\theta)$ is unitarily
equivalent to
\begin{equation}
  {H}_{0}(\theta)=(-i\partial_x +By+\theta )^2 -\partial_y^2 +\omega^2 y^2
\quad {\rm on}\quad L^2([0,2\pi]\times\R)
\end{equation}
with periodic boundary conditions at $x=0$ and $x=2\pi$.
We exhibit below a complete set of eigenvectors in 
\be
D_e\equiv\{f\in W^{2,2}([0,2\pi]) | f(0)=f(2\pi), f'(0)=f'(2\pi)  \}
\otimes S(\R)
\ee
where $S(\R)$ denotes the set of Schwarz function,
showing that $H_0(\theta)$ is essentially self adjoint on $D_e$.
Next we show that $H_0(\theta)$ is a  holomorphic family of type A
in the sense of Kato. Let $H_0(0)$ is self-adjoint on its
domain $D$ and let us formally expand the operator $H_0(\theta)$
as
\begin{equation}\label{expansion}
  H_0(\theta)=(-i\partial_x +By)^2 -\partial_y^2 +\omega\, y^2
+2 \theta(-i\partial_x +By) +\theta^2.
\end{equation}
We note that $(-i\partial_x +By)$ is symmetric on $D_e$ and
denote the resolvent by $R_0(\theta,z)=(H_0(\theta)-z)^{-1}$.
Now, for any $\ffi\in D_e$ 
\bea
&& \|(-i\partial_x +By)\ffi\|^2 \leq \bra \ffi | H_0(0) \ffi\ket
=\bra R_0(0,z)(H_0(0)-z)\ffi | H_0(0) \ffi\ket\nonumber\\
&& \leq
\| R_0(0,z)\| \|H_0(0)\ffi\|^2+|z|\bra \ffi | R_0(0,\bar{z})H_0(0)\ffi\ket\nonumber\\
&& \leq C(z)  \|H_0(0)\ffi\|^2 +|z|^2 C(z) \|\ffi\|^2,
\label{positive}
\eea
where $C(z)=O(1/\Im z)$, as $\Im z \ra \infty$, $|\Re z| < \infty $.
From Theorem V.4.4 p.288 in \cite{kato}, we deduce that 
that $(-i\partial_x +By)$ is relatively bounded with respect to
$H_0(0)$ on $D_e$, with arbitrarily small relative bound (to this end, take
$|\Im z |$ large enough). Hence the domain of $H_0(\theta)$
coincides with $D$ for any complex $\theta$, and
the expansion (\ref{expansion}) shows that the vector
$H_0(\theta)\psi$ is holomorphic in $\theta$ for any $\psi\in D$.
That means $H_0(\theta)$ is a self-adjoint holomorphic family of
type A in the whole complex plane, see~\cite{kato}, pp.~375 and 385.
The same is true for the perturbed operator
\be
H(\theta)=H_0(\theta)+W(x,y)
\ee
when $W$ is bounded.

In order to find the spectrum of $H_0(\theta)$ we introduce the basis
\be
\psi_m(x)=(2\pi)^{-1/2} \exp(i\, m x) \label{bas}
\ee
and get the decomposition
\bea
 {H}_{0}(\theta)&=&\bigoplus_{m\in {\bf Z}}\, |\psi_m
\ket H_0^m(\theta)\bra \psi_m| \\ &=&\bigoplus_{m\in {\bf Z}}\, |\psi_m
\ket \left [( m +By+\theta )^2 -\partial_y^2 +\omega^2 y^2\right
]\bra \psi_m|,\nonumber
\eea
where $H_0^m(\theta)= \bra \psi_m| H_0(\theta) \psi_m \ket$.
By a
unitary transform inducing a $(\theta+m)$-dependent shift of the
argument we find that $H_0^m(\theta)$ is unitarily equivalent to
\be
  h_m(\theta)=-\partial_u^2+\alpha^2u^2+\beta(m+\theta)^2,
\ee
with $\alpha=\sqrt{B^2+\omega^2}$, $\beta=\omega^2/
(B^2+\omega^2)$, and $u=y+B(m+\theta)/(B^2+\omega^2)$. This
operator is clearly analytic in $\theta$. Therefore we get the
spectrum
\be
\sigma (H_0^m(\theta))=
\left\{\alpha(2n+1)+\beta(m+\theta)^2\right\}
=\left\{E_n(\theta+m)\right\}_{n\in{\bf N}_0}\, ,
\label{spect} 
\ee
where the corresponding eigenfunctions of $H_0^m(\theta)$,
$\ffi_n^{m+\theta}(y)$, are translates of the usual harmonic
oscillator states $\ffi_n(u)$. More precisely, if $V_{\theta+m}$
is the unitary operator from $L^2({\R}_y)$ to  $L^2({\R}_u)$
defined by
\be
(V_{\theta+m}f)(u)=f(u-B(m+\theta)/(B^2+\omega^2)),
\ee
then $V_{\theta+m}H^m_0 V_{\theta+m}^{-1}=h_m(\theta)$ and
$\ffi_n^{m+\theta}(y)=(V_{\theta+m}^{-1}\ffi_n)(y)$. For a later
purpose, let us also introduce the unitary operator $V(\theta)$
from $L^2({\R}_x\times{\R}_u)$ to $L^2({\R}_x\times{\R}_y)$ given
as
\be
  V(\theta)=\bigoplus_{m\in{\bf Z}}V_{\theta+m}.
\ee
Let us turn to
\be
H(\theta) = H_0(\theta)+W(x,y) \quad {\rm on}\quad L^2([0,2\pi]\times\R)
\ee
with periodic boundary conditions at $x=0$ and $x=2\pi$. Since
$W$ is bounded, it is relatively compact w.r.t. $H_0(\theta)$
and the essential spectrum of $H(\theta)$ is thus the same as that
of $H_0(\theta)$. It follows that $\sigma (H(\theta))$ is
discrete. The corresponding eigenvalues are analytic functions of
$\theta$, we denote them as $E_j(\theta)$.

At this point, we see that for any $E'>0$, and uniformly
in $|\theta|<1/2$, there are finitely many eigenvalues
of $H_0(\theta)$ $E_{n,m}(\theta)=\alpha(2n+1)+\beta(m+\theta)^2$ 
below $E'$. These eigenvalues being are branches of analytic 
functions in $\theta$  may display finitely many crossings 
with one another. The same 
is true for those of the perturbed operator $H(\theta)$. 
In order to exclude the possibility for a perturbed
eigenvalue to be constant in $\theta$, it is enough to impose 
that the perturbation be smaller than half the smallest variation
of the finitely many arcs of analytic functions 
free from crossings below $E$.  
Therefore, below $E=E'-\| W \|$, the eigenvalues
of $H(\theta)$ cannot be constant and we have
\begin{prop}
For any $E>0$, the spectrum of the Hamiltonian
$(\ref{depart})$ is purely absolutely continuous below $E$
if $\| W\|_{\infty}$ is small enough.
\end{prop}

Let us turn to the case where $W$ depends on $x$ only.
We are interested in the properties of the eigenvalues of
$H_0^m(\theta)$, which coincide with those of $h_m(\theta)$. As
the eigenfunctions of $h_m(\theta)$ are independent of $m+\theta$,
it is easier to deal with this operator as $\theta$ becomes
complex than with $H_0^m(\theta)$. We define
\be
h_0(\theta)=V(\theta)\, H_0(\theta)\, V^{-1}(\theta)\,,
 \ee
then we have the relation
\be
\|(h_0(\theta)+1)^{-1}\|^2= \sup_{m\in {\bf
Z}}\|r_m(\theta)r_m(\theta)^*\|,\quad
r_m(\theta):=(h_m(\theta)+1)^{-1}  
\ee
When $\theta$ becomes complex, in which case we will write
$\theta=\theta_1+i\, \theta_2$, the resolvent
$r_m(\theta)$ remains compact and
$r_m(\theta)^* =(h_m(\overline{\theta})+1)^{-1}$ so that
\be
 \|r_m(\theta)r_m(\theta)^*\|=\sup_{n\in {\bf N}_0}
 \frac{1}{|E_n(\theta +m)+1|^2}\,,
\ee
since the basis $\left\{\ffi_n(u)\right\}_{n\in{\bf N}_0}$ remains
orthonormal for complex $\theta$. Then one can show that this norm
goes to zero as $\theta\ra\infty$ in some direction of the complex
plane, uniformly in $m \in \mathbb{Z}$. Indeed, from (\ref{spect})
we get
\bea \lefteqn{
 \|r_m(\theta)r_m(\theta)^*\|} \nonumber \\ && = \sup_{n\in {\bf
     N}_0}\frac{1}{[\alpha(2n+1)+\beta((m+\theta_1)^2
     -\theta_2^2)+1]^2+[2\beta\theta_2(m+\theta_1)]^2} \nonumber \\
&& \leq \frac{1}{[2\beta\theta_2(m+\theta_1)]^2}\,, \eea
which goes to zero as $\theta_2\ra\infty$ uniformly in $m$
provided $\theta_1$ is not an integer.

Furthermore, from the fact that $h_0(\theta)$ is a self-adjoint
holomorphic family of type A it follows that
$(h_0(\theta)+1)^{-1}$ is compact either for all $\theta$ or for
no $\theta$ -- cf. \cite[Thm.~VII.2.4]{kato}. We have already seen
that $(h_0(\theta)+1)^{-1}$ is compact for $\theta$ real, 
so it is compact also for $\theta$ complex. Thus
$(h_0(\theta)+1)^{-1}\, (h^*_0(\theta)+1)^{-1}$ is a compact
self-adjoint operator, and since the family
$\left\{\ffi_n(u)\right\}_{n\in{\bf N}_0}$ still forms a complete
orthonormal basis in  $L^2( \R)$, the eigenvalues of $h_0(\theta)$
retain the form (\ref{spect}). Hence one has
\be
\|(h_0(\theta)+1)^{-1} (h_0(\overline\theta)^*+1)^{-1}\|
=\|(h_0(\theta)+1)^{-1}\|^2\leq \frac{1}{\beta^2\theta_2^2}\,,
\label{ubound} \ee
where we have chosen for simplicity $\theta_1=1/2$. 

The perturbed fiber operator is
\be
h(\theta)= h_0(\theta)+V(\theta)\, W(x)\,
V^{-1}(\theta)=h_0(\theta)+W(x) 
\ee
The point is now to show, that the eigenvalues $E_j(\theta)$ of 
$h(\theta)$ are not constant in $\theta$. Then the same is true, 
for $\theta$ real, also for the eigenvalues of
\be
H(\theta) = H_0(\theta)+W(x,y)
\ee
and this yields the absolute continuity of (\ref{hamiltonian}). 

We use Thomas argument -- \cite{tom} and
\cite[Sec.~XIII.16]{rees} -- and assume that some $E_j(\theta)$ is
equal to $E_0$ for all $\theta$. From the above analysis it
follows that $E_0$ is an eigenvalue of $h(\theta)$ also for all
complex $\theta$, and therefore
\be
\|(h(\theta)+1)^{-1}\|\geq (E_0+1)^{-1} \label{lowerbound}
\ee
On the other hand, a standard argument based on the resolvent
identity shows that  for $\|W(x)
(h_0(\theta)+1)^{-1}\|<1$ (i.e. $\theta_2$ large enough
-- cf. (\ref{ubound})) is
\be
\|(h(\theta)+1)^{-1}\|\leq\frac{\| (h_0(\theta)+1)^{-1}\|}{1-\|W(x)
(h_0(\theta)+1)^{-1}\|}
\ee
so $\|(h(\theta)+1)^{-1}\|\ra 0$ as $ \theta_2\ra\infty$ by
(\ref{ubound}). In this way we get a contradiction with
(\ref{lowerbound}), so no $E_j(\cdot)$ can be constant.

Finally, we note also that if $W(x,y)\equiv W(y)$ is bounded and depends 
on $y$ only, we get by simple manipulations that $H$ is unitarily 
equivalent to
\be
H \simeq \int_{p\in {R}}^{\oplus} H(p)\, \D p
\ee
where 
\be
H(p) =  -\pd_y^2+\alpha^2 y^2 +p^2 \frac{\omega^2}{B^2+\omega^2} + 
W(y- p\, B/(B^2+\omega^2)).
\ee
As $W$ is bounded, we see that the analytic eigenvalues 
$\{e_n(p)\}_{n\in {\bf N}}$ of $H(p)$ tend to $\alpha(2n+1)+
p^2 \frac{\omega^2}{B^2+\omega^2}$ as $p\rightarrow \infty $.
Therefore they cannot be constant and the spectrum of $H$ 
is purely absolutely continuous also.

This allows us
to make the following claim
\vspace{2mm}
\begin{thm} \label{thomas}
Let $W_1(x)\in L^{\infty}(\R)$ be periodic in $x$ and 
$W_2(y)\in L^{\infty}(\R)$. Then the spectra of
both operators 
\bea
H &=&  -\pd_y^2+(-i\pd_x+yB)^2+\omega^2 y^2 + W_1(x)\\
H &=&  -\pd_y^2+(-i\pd_x+yB)^2+\omega^2 y^2 + W_2(y)
\eea
are purely absolutely continuous for any $\omega\ne 0$.
\end{thm}

\section{Open gaps} \label{opengaps}

The result of previous section shows that the absolute
continuity of the bottom of the spectrum of the magnetic Hamiltonian in the
presence of a parabolic confinement is not affected by a 
small bounded $x$-periodic perturbation. Of course, one would like 
to know how the
spectrum looks like as a set, in particular how many gaps can open
as a consequence the perturbation. We now show that for a
non-constant $W(\cdot,y)$ there are generically many gaps in the
spectrum of $H$ provided the coupling constant of the confinement
is large enough.

We start again with the fiber Hamiltonian
\be
H(\theta) = -\pd_y^2+(-i\pd_x+yB)^2+\omega^2 y^2 + W(x,y) \label{h0}
\ee
on $L^2([0,2\pi]\times\R)$ with the boundary conditions
(\ref{bc}). We introduce a new variable $s$ by
\be
s =\sqrt{\alpha}\, y,\quad  \alpha:= \sqrt{B^2+\omega^2}
\ee
and the orthonormal basis on $L^2(\R)$
 \be
 \varphi_n(s) = C_n\exp(-s^2/2)\,
H_n(s),\quad C_n=(1/\pi)^{1/4}\, (2^n n!)^{-1/2},\quad n\in {\bf N_0}
\label{basis}
 \ee
Let us introduce some more notations,
\bea
W^{(\alpha)}_{n,m}(x) & = & (\ffi_n,W
\ffi_m)=\int_{\R}\ffi_n(s)\, \ffi_m(s)W(x,s/\sqrt{\alpha})\, \D\, s,\, n\neq
m \nonumber \\ 
W^{(\alpha)}_n(x) & = & (\ffi_n,W
\ffi_n)=\int_{\R}\, \ffi_n(s)\ffi_n(s)W(x,s/\sqrt{\alpha})\, \D\, s
\eea
The matrix elements of $H(\theta)$ in the basis (\ref{basis}) are
then the operators on $L^2([0,2\pi])$ given by
\bea \label{matrix}
 H_{n,m}(\theta) & = &
\delta_{n,m}\left [\alpha(2n+1)+K_n(\theta)\right ] +
W^{(\alpha)}_{n,m}(x)(1-\delta_{n,m}) \nonumber \\
 & - &
 \delta_{n+1,m}\sqrt{\frac{2(n+1)}{\alpha}}\, i\,
   B\pd_x-\delta_{n-1,m}\sqrt{\frac{2n}{\alpha}}\, i\, B\pd_x\,,
\eea
where we define $K_n(\theta)$ as
\be
K_n(\theta) = -\pd_x^2 + W^{(\alpha)}_n(x) \ee
with the domain
$$ D(\theta)=\left\{f\in W_{2,2}[0,2\pi];\, f(2\pi)=e^{2\pi
  i\theta}f(0),f'(2\pi)=e^{2\pi i\theta}f'(0)\right \}
$$
By \cite[Sec.~XIII.16]{rees} for each $n\in {\bf N_0}$ the
operator $K_n(\theta)$ has a purely discrete spectrum,  and none of
their eigenvalues is constant in $\theta$. We will
denote the eigenvalues and eigenfunctions of $K_n(\theta)$ by
\be
\eps_k(n,\theta);\, \psi_k^n(x,\theta),\quad k\in \mathbb{Z}\,,
\ee
respectively, where for any fixed $\theta$ and $n$ the functions
$\psi_k^n(x,\theta)$ form an orthonormal basis in $L^2[0,2\pi]$.
It is shown in \cite[Thm.~XIII.91]{rees} that for a non-constant
$W_n$ at least one gap is present in the spectrum of
$$ K_n:= \int_{|\theta|\le 1/2}^{\oplus} K_n(\theta)\, \D\theta
\label{direct} $$
In other words, there exists some $j$ such that
\be
 \sup_{|\theta|\leq 1/2} \eps_j(n,\theta) < \inf_{|\theta|\leq
 1/2} \eps_{j+1}(n,\theta)
\ee
We are particularly interested in the spectrum of $H_{0,0}$, the
direct integral from $H_{0,0}(\theta)$ over $\theta$, which
contains at least one gap if $W^{(\alpha)}_0$ is not constant.

It follows from (\ref{matrix}) that taking $\alpha$ large enough,
this gap will not be covered by the spectra of the other diagonal
elements of $H_{n,m}(\theta)$. Then one needs only show that this
gap remains open after taking into account the off-diagonal
elements of $H_{n,m}(\theta)$. To see that, we apply 
perturbation theory. As unperturbed operator we take
\be
H^D(\theta) = \bigoplus_{n\in {\bf N_0}}\, H_{n,n}(\theta)\quad {\rm
  on}\quad L^2[0,2\pi]\times l_2
\ee
with eigenvalues and eigenvectors given by
\be
\alpha (2n+1)+\eps_k(n,\theta)\,,\quad  \psi_k^n(x,\theta)\left
(\matrix{0\cr 0\cr 1\cr 0\cr \vdots \cr}\right ) \ee
respectively, where $1$ stands in the $n$-th row. 
Moreover, we have

\vspace{0.1cm}

\begin{lem} 
Let $H^{OD}(\theta) = H(\theta)-H^D(\theta)$. Then
\be
\|H^{OD}(\theta)(H^D(\theta)+i)^{-1}\|=\mathcal{O}(1/\alpha), \quad {\rm as}\quad
\alpha\ra\infty 
\label{relativeb}
\ee
uniformly in $\theta$.
\end{lem}

\vspace{0.2cm}

{\it Proof:} For 
$$
W^D=\bigoplus_{n\in {\bf N_0}}\, W^{\alpha}_n(x)
$$ 
we define $W^{OD}=W-W^D$. Then
\be
\|W^{OD}(H^D(\theta)+i)^{-1}\|\leq
2\|W\|_{\infty}\|(H^D(\theta)+i)^{-1}\|=\mathcal{O}(1/\alpha) 
\label{wbound}
\ee
as $\alpha\ra\infty$ since ${\rm dist}(\sigma(H^D(\theta)),i)$ grows linearly
with $\alpha$. 

Let us now take $n$ fixed. For the other elements of $H^{OD}(\theta)$, i.e. the
last two terms on the r.h.s. of (\ref{matrix}), we have
\be
\left (i\pd_x\pm \sqrt{\alpha(2n+1)}\right )^2>0,\quad \pm\, 
2i\sqrt{\alpha(2n+1)}\, \pd_x\leq -\pd_x^2+\alpha(2n+1)
\ee
so that as quadratic forms on $D(\theta)$
\be
-\frac{B^2}{\alpha}\, 2(n+1)\, \pd_x^2\leq\, \frac{B^2}{\alpha^2}\,
(-\pd_x^2+\alpha(2n+1))^2 
\ee  
Then, in the sense of
(\ref{relativeb}),    
\bea
& & \| |\psi_n\ket \bra \psi_n | iB\alpha^{-1/2}\sqrt{2(n+1)}\, 
\pd_x\, |\psi_{n+1}\ket \bra \psi_{n+1} |(H^D(\theta)+i)^{-1}\| 
\nonumber \\
& & =\|  iB\alpha^{-1/2}\sqrt{2(n+1)}\, 
\pd_x\, (H_{n+1, n+1}(\theta)+i)^{-1}\| 
\nonumber \\
& & \leq 
\frac{B}{\alpha}\|(-\pd_x^2+\alpha(2n+1))
(-\pd_x^2+W^{\alpha}_{n+1}\alpha(2n+3)+i)^{-1}\| \nonumber \\
& \leq &
\frac{B}{\alpha}\,
\left({1+\|W^{\alpha}_{n+1}(-\pd_x^2+W^{\alpha}_{n+1}+\alpha(2n+3)+i)^{-1}\|}
\right)
=\mathcal{O}(1/\alpha) 
\eea
as $\alpha\ra\infty$,  uniformly in $n$. Inequality (\ref{wbound}) and
the Schur condition, 
\cite[Ex.~III.2.3]{kato}, then give the statement of the Lemma. \QED

\vspace{0.5cm}

Now the resolvent identity in combination with (\ref{relativeb}) implies
\bea
& & \|(H(\theta)+i)^{-1}-(H^D(\theta)+i)^{-1}\| = \nonumber \\ 
& = & \|(H(\theta)+i)^{-1} H^{OD}(\theta)(H^D(\theta)+i)^{-1} \|\, 
\ra 0\quad {\rm as}\quad \alpha\ra\infty
\eea
so that $H^D(\theta)$ converges to $H(\theta)$ in norm resolvent
sense, uniformly in $\theta$. From perturbation theory,
see \cite[Thm.~IV.2.25]{kato}, we thus get the convergence of
spectra of $H^D(\theta)$ and $H(\theta)$. 
It follows
that for large enough $\alpha$, keeping $B$ fixed, the gap between
$\eps_j(0,\theta)$ and $\eps_{j+1}(0,\theta)$ will be open also in
the spectrum of $H$. The argument works for any fixed
$j\in\mathbb{Z}$, i.e. sending $\alpha\ra\infty$ we can keep any
finite family of gaps contained in $\sigma(H_{0,0})$ open. We have
thus proven

\vspace{2mm}
\begin{thm} \label{gaps}
Let $W(x, y)\in L^{\infty}(\R^2)$. Denote by $N(H)$ and
$N(H_{0,0})$ the number of open gaps in the spectrum of $H$ and
$H_{0,0}$ respectively. If $N(H_{0,0})$ is finite, then $N=
N(H_{0,0})$ holds for $\omega$ large enough; in particular, an
open gap exists for a sufficiently strong confinement whenever the
function $W_0$ is non-constant. If $N(H_{0,0})=\infty$, then to
any positive integer $n$ there is $\omega(n)$ such that
$$ N(H)\geq n $$
holds for all $\omega\geq\omega(n)$.
\end{thm}

\vspace{2mm}

\noindent {\bf Remark:} It is also clear from the above given argument, that
taking $\omega$ large enough gives us the absolute continuity of
$\sigma(H)$ in the bottom of the spectrum. More precisely, in the
interval $\left [\, \inf \sigma(H_{0,0}),\, \inf \sigma(H_{1,1})\right ]$.

\section{Transport in presence of localized \\ perturbations} \label{scatt}

As we have indicated in the introduction, we turn now to
situations when the perturbation is not periodic, but bounded and
localized in a sense to be precised below. In this case we have
\be
H = H_0 + W = -\pd_y^2+(-i\pd_x+yB)^2+\omega^2 y^2 + W(x,y) \label{ran}
\quad {\rm on} \quad L^2(\R^2)
\ee
with $W(x,y)\in L^{\infty}(\R^2)$. By \cite[Chap.~X]{rees} the
Hamiltonian (\ref{ran}) is e.s.a. on $C_0^{\infty}(\R^2)$. For later
purposes we notice that $S(\R^2)$, the Schwarz functions, is also a
core for $H$. This follows from the fact, that $H$ is clearly
symmetric on $S(\R^2)$ and $C_0^{\infty}(\R^2)$ is included in
$S(\R^2)$.
The question is the following: in what part of the spectrum and under
which conditions does transport survive in the presence of the
impurity potential $W(x,y)$?

Instead of the Bloch decomposition we now employ the 
commutator method. The point is to find a suitable
conjugate operator $A$ which satisfies the Mourre estimate
\be
E_{\Delta}(H)[H,iA]E_{\Delta}(H) \geq \kappa\, E_{\Delta}(H)
\label{mest} \ee
for some strictly positive constant $\kappa$. Here $E_{\Delta}(H)$
is the spectral projection of $H$ on the interval $\Delta$. 
Then, under some regularity assumptions on $H$, we can obtain the
absence of point spectrum in the interval $\Delta$ using the Virial
Theorem, \cite{geg} 

\vspace{2mm}

\begin{thm}[Virial] \label{virial}
Let $H,A$ be self-adjoint operators on $L^2(\R^2)$ and assume that
$H$ is of class $C^1(A)$, i.e. there is  $z\in\rho (H)$ such that  
\be 
\R\ni t\mapsto e^{itA}(z-H)^{-1}e^{-itA}
\label{map}
\ee 
is of class $C^1$ in the strong operator topology. Then
$$
(\psi,[H,iA]\psi)=0
$$
for any eigenfunction $\psi$ of $H$.
\end{thm}

Under stronger hypothesis on $H$, we can apply the
Mourre theorem -- cf.~\cite{mourre},\cite{abg} -- and exclude even the
possibity of singular continuous spectrum in $\Delta$.
For a precise statement of the Mourre Theorem, we 
have the formulation from \cite{sah1, sah2}. 

\vspace{2mm}
\begin{thm}[Mourre] \label{mourre}
Let $H,A$ be self-adjoint operators on $L^2(\R^2)$ and assume that
\begin{enumerate}
\item There is $\alpha>0$ such that $H$
is of class $C^{1+\alpha}(A)$, i.e. $H$ is $C^1(A)$ and the derivative
of $(\ref{map})$ is H\"older continuous of order $\alpha$.
\item $H$ and $A$ satisfy the estimate $(\ref{mest})$ for an open
  interval $\Delta$ and $\kappa>0$.
\end{enumerate}
Then the spectrum of $H$ in the interval $\Delta$ is purely absolutely
continuous.
\end{thm}

\noindent {\bf Remark:} We shall use the last theorem with $\alpha=1$, which
corresponds to the original formulation given in \cite{mourre}, see also 
\cite[Thm.~4.9]{cycon}.
\par
The classical counterpart of the positive commutator (\ref{mest})
is an observable which increases in time. To find a suitable
candidate for the conjugate operator in our case, let us therefore
discuss first the classical dynamics of the unperturbed system.

\subsection{Classical solution in the absence of perturbation}

We will denote the position vector of the particle by
$(x(t),y(t))$. In the absence of $W(x,y)$ the classical
Hamiltonian is of the form
\be 
H_{cl}=(p_x+yB)^2+p_y^2+\omega^2\, y^2
\ee
where
\be
p_x(t)=\frac{1}{2}\, \dot x(t)-y(t)\, B,\quad p_y(t)
=\frac{1}{2}\, \dot y(t)
\ee
From  Hamilton's equations we thus get
\be
\dot p_x(t)=0,\quad \dot
p_y(t)= -\dot x(t)B-2\omega^2\, y(t)
\label{hamilton}
\ee
Given initial conditions $x(0),y(0),p_x(0),p_y(0)$, the solution
of (\ref{hamilton}) reads
\bea x(t) & = & -\frac{B}{2\alpha^2}\, p_y(0)\cos (2\alpha
t)+\frac{B}{\alpha}\left (y(0)+\frac{B}{4\alpha^2}\, p_x(0)\right
)\sin (2\alpha t) \nonumber \\ && +\, 2p_x(0)\, t\,
\frac{\omega^2}{\alpha^2}+ x(0)+\frac{B}{2\alpha^2}\, p_y(0)
\nonumber \\ y(t) & = & (2\alpha)^{-1}p_y(0)\sin (2\alpha t)+ \left
  (y(0)+\frac{B}{4\alpha^2}\, p_x(0)\right )\cos (2\alpha
t)-\frac{B}{\alpha^2}p_x(0) \nonumber \\ p_x(t) & = & p_x(0)
\nonumber \\ p_y(t) & = & \frac{1}{2}\, p_y(0)\cos (2\alpha t)
-\, \alpha\left
  (y(0)+\frac{B}{4\alpha^2}\, p_x(0)\right )\sin(2\alpha t)
\label{solution}
\eea
Note that the momentum $p_x$ is preserved since the free
Hamiltonian $H_0$ commutes with $x$-translations, see
(\ref{hamilton}). It is easy to 
see that the classical trajectory is now given by an ellipse, with
the position vector of its center being
\be
S(t) = \left [2p_x(0)\, t\,
\frac{\omega^2}{\alpha^2}+x(0)+\frac{B}{2\alpha^2}\,
p_y(0),-\frac{B}{\alpha^2}p_x(0)\right ]\,, \ee
so that as long as $\omega\neq 0$, i.e. the confinement is present,
the center of the ellipse is moving along the $x$ axis with the
constant velocity and in the direction given by a sign of the initial
momentum $p_x(0)$. Note also, that the two ellipses which correspond
to the motions in opposite directions are mutually shifted by
$\frac{2B}{\alpha^2}p_x(0)$.

A classical observable whose absolute value is increasing in time
is thus the $x-$  component of $S(t)$, which can be written as
\be
S_x(t) = x(t)+ \frac{B}{\alpha^2}\,p_y(t)\,. \label{center} \ee
However, since we need something which has a definite sign
independently of the initial conditions, we multiply
(\ref{center}) by $p_x(t)$; then
\be
\pd_t(p_x(t)S_x(t)) = 2\, p^2_x(0)\frac{\omega^2}{\alpha^2}>0\,.
\ee
In other words, the corresponding quantum mechanical conjugate
operator can be chosen in the form
\be
A = \frac{1}{2}(-i\pd_x\, x-ix\, \pd_x)-\frac{B}{\alpha^2}\,
\pd_x\pd_y\,. \label{conj} \ee

\subsection{Absence of eigenvalues and absolute continuity}

Now we are going to show that under some regularity and decay
assumptions on $W$ the absolutely continuous spectrum of the free
Hamiltonian persists in some parts of the spectrum of $H$. In
particular, this makes scattering on the impurity in our
parabolic channel possible.

The conditions we impose on $W(x,y)$ then are as follows:

\begin{itemize}
\item[$(a)$]  $W_0 \!:=\! \|W\|_{\infty}<\alpha,\, 
W'_0:=\|x\, \pd_xW\|_{\infty}<\infty $

\item[$(b)$]  $W\in C^2(\R^2)$ and 
$$
\|\pd_x^2W\|_{\infty}<\infty,\,
  \|\pd_y^2W\|_{\infty}<\infty,\, \|\pd_x\pd_yW\|_{\infty}<\infty,\,
  \|x^2\pd_x^2W\|_{\infty}<\infty
$$   
\end{itemize}

Before looking for the Mourre estimate, we check the
regulariry of the map (\ref{map}).
 
First we state an auxiliary Lemma, which is proven in the Appendix.

\vspace{2mm}

\begin{lem} \label{relbound}
There exists a number $c$ such that
\begin{itemize}
\item[$(i)$] $\|\pd_y^2\, R_0(\lambda)\| \leq c$
\item[$(ii)$]
$
\|\pd_x^2\, R_0(\lambda)\|,\, 2\, \|y\, \pd_x\,
R_0(\lambda)\|,\,\|y^2\, R_0(\lambda)\| \leq c\,
\frac{1+\alpha^2}{\omega^2}
$
\item[$(iii)$] $\|\pd_x\pd_y\, R_0(\lambda)\|\leq c\sqrt{\frac{1+\alpha^2}{\omega^2}}$
\end{itemize}
where $R_0(\lambda)=(H_0+\lambda)^{-1},\quad  \lambda\geq 0$.
\end{lem}
  
\vspace{2mm}

\noindent Now we show that under the assumption $(a)$ one can apply the
Virial Theorem to a pair of operators $H,\, A$. 

\begin{lem} \label{c1}
Let $W(x,y)$ satisfy the condition $(a)$. Then $H$ is of class $C^{1}(A)$.
\end{lem}

{\it Proof:} By \cite{geg} and \cite[Thm.~6.3.4]{abg} to show that $H$ 
is $C^1(A)$, it is enough to prove that  
\begin{itemize}
\item[$(1)$] $e^{itA}$ preserves $D(H)$,

\item[$(2)$] There is a constant $c$ such that
$$
|(H\ffi,A\ffi)-(A\ffi,H\ffi)|\leq c\, (\|H\ffi\|^2+\|\ffi\|^2),\,
\ffi\in D(H)\, \cap\, D(A).
$$
\end{itemize}

Since $W$ is bounded, the domain of $H$ coincides
with that of $H_0$ and we can thus check the condition $(1)$ only for
$D(H_0)$. Let $D$ be a core for $H_0$. It follows from
\cite[Lem.~7.6.5]{abg}, that to prove $(1)$ it suffices to show, 
in addition to $(2)$, that
\begin{itemize}
\item[(i)] for $u\in D$ and $t\in\R$, $e^{itA}u\in D$
and $\sup_{|t|\leq 1}\|H_0e^{itA}u\|<\infty$.
\item[(ii)] the derivative $\pd_te^{-itA}H_0e^{itA}u|_{t=0}\equiv
[H_0,iA]u$ exists weakly for each vector $u\in D$.
\end{itemize}
To begin with, we notice that $A$ being quadratic in momentum
and position, we know by \cite[Thm.~3.4]{hag} that the 
unitary propagator $U(t)=e^{-itA}$ is such that
$$
U(t): S(\R^2)\mapsto S(\R^2)
$$
Now, $S(\R^2)$ is a core for $H_0$, so the first part of (i) is
satisfied. To see how $U(t)$ acts on the function from $S(\R^2)$,
we apply a partial Fourier
transformation in $y$, and denote the transformed operators by
$\widehat{H_0}$ and $\widehat{A}$. It can be directly checked, that for
any $\psi(x,y)\in S(\R^2)$ 
\be
e^{-it\widehat{A}}\widehat{\psi}(x,k)  = e^{-t/2}
  \widehat{\psi}\left (e^{-t}x-k\mu(1-e^{-t}),k\right )
\label{transf}
\ee
where $\widehat{\psi}(x,k)={\cal F}_y\, \psi(x,y)$ and $\mu:=\frac{B}{\alpha^2}$.

A simple calculation then gives 
\bea\label{simple}
& & e^{-it\widehat{A}}\, \widehat{H}_0\, e^{it\widehat{A}}\widehat{\psi}(x,k)= \nonumber \\
& = & a(t)\, \pd_x^2\, \widehat{\psi}(x,k)+b(t)\pd_x\pd_k\widehat{\psi}(x,k)
-\alpha^2\pd_k^2\, \widehat{\psi}(x,k)
+k^2\widehat{\psi}(x,k)
\eea
where
\bea
a(t) & = & -e^{2t}\left (1+2Be^t\mu
  (1-e^t)+\alpha^2e^{2t}\mu^2(1-e^t)^2\right) \nonumber \\
b(t) & = & -e^t\left (2B+2\alpha^2e^t\mu (1-e^t)\right )
\eea
are both $C^{\infty}$, so that the second
part of $(i)$  and $(ii)$ hold.


Moreover, it is easily seen from (\ref{transf}) that $U(t)$ is
strongly differentiable on $S(\R^2)$. 
It follows then from \cite[Thm.~VIII.10]{rees} that $A$ is
essentialy self-adjoint on $S(\R^2)$.    

This allows us to verify the
condition $(2)$ only on functions in $S(\R^2)$. First we notice that $H$ can
be written as
\be
H =
\left (-i\pd_x\frac{B}{\alpha}+y\alpha\right
)^2-\beta\, \pd_x^2- \pd_y^2 + W(x,y)
\ee
reminding that
$$
\beta=\frac{\omega^2}{\alpha^2}
$$
Then for any $\ffi\in S(\R^2)$
\bea
& & |(H\ffi, A\ffi)-(A\ffi, H\ffi)| \leq |(\ffi,-2\, \beta\,
\pd_x^2\, \ffi)| \nonumber \\
& & +
\mu|(W\ffi,\pd_x\pd_y\ffi)-(W\ffi,\pd_x\pd_y\ffi)| +|(\ffi,(\pd_x W)x\ffi)| \nonumber \\
& & \leq 2|(\ffi,H_0\ffi)| +
2\mu W_0 \|\ffi\|\, \|\pd_x\pd_y\ffi\|+\|\ffi\|^2W'_0  
\label{comm} 
\eea
On the other hand we have
\bea
\|i\pd_x\ffi\|^2 & \leq & \beta^{-1}\, \|\ffi\|\, \|H_0\ffi\|\leq
\beta^{-1}\, \|\ffi\|(\|H\ffi\|+W_0\|\ffi\|) \nonumber \\ 
\|i\pd_y\ffi\|^2 & \leq & \, \|\ffi\|\, \|H_0\ffi\|\leq
\|\ffi\|(\|H\ffi\|+W_0\|\ffi\|) 
\eea
and since  $H\geq\alpha-W_0>0$ holds by assumption, also
\be
\|\ffi\|\leq (\alpha-W_0)^{-1}\|H\ffi\|
\ee
Moreover, it follows from Lemma \ref{relbound}, that 
\be
\|\pd_x\pd_y\ffi\| \leq {\rm const}\, \|H_0\ffi\|
\ee
Using all the inequalities we can find some large
enough constant $c$, depending on $\alpha$ and $W_0$, such that
\be
|(H\ffi, A\ffi)-(A\ffi, H\ffi)|\leq c\, (\|H\ffi\|^2+\|\ffi\|^2)
\ee
proving thus $(2)$. 

Finally, $(2)$ in combination with
\cite[Lem.~7.6.5]{abg} shows that $e^{it\widehat{A}}$ preserves
$D(\widehat{H_0})$. That is,
for any $\psi(x,y)\in D(H_0)$ we have $e^{it\widehat{A}}\widehat{\psi}(x,k)\in
D(\widehat{H}_0)$ and
 \be
 e^{itA}\psi(x,y)={\cal F}_y^{-1}\,
 e^{it\widehat{A}}\widehat{\psi}(x,k)\in {\cal F}_y^{-1}\, 
 D(\widehat{H}_0)=D(H_0) \label{pres}
 \ee
which completes the proof of the Lemma. \quad \QED

\vspace{2mm}

The hypothesis of the Mourre theorem require a slightly stronger regularity of
$H$. We will impose some additional assumptions on $W(x,y)$. 

\vspace{2mm}
 
\begin{lem} \label{c1+a}
Assume $(a)$ and $(b)$. Then $H$ is $C^{2}(A)$. 
\end{lem}

\noindent
{\it Proof:} We will prove the statement of the Lemma separately for
$H_0$ and $W$.
\\
First we prove that $H_0$ is $C^{\infty}(A)$. We work in the Fourier
picture, as above. Consider
\be
\widehat{H}_0(t)=e^{-it\widehat{A}}\, \widehat{H}_0\, e^{it\widehat{A}},
\ee
self adjoint on $D(\widehat{H_0})$ for any $t\in\R$ and, for $\lambda>\|W\|+1$,
\be
\widehat{R}_0(t)=e^{-it\widehat{A}}\,(\widehat{H_0}+\lambda)^{-1}e^{it\widehat{A}}.
\ee
As $\widehat{R}_0(t+t_0)=e^{-it_0\widehat{A}}\,\widehat{R}_0(t)e^{it_0\widehat{A}}$,
it is enough to check differentiability at $0$.
From the resolvent identity on $(\widehat{H}_0+1)S(\R^2)$ and
(\ref{simple}), we get
\bea
& & \widehat{R}_0(t)-\widehat{R}_0(0)=-\widehat{R}_0(t)(\widehat{H}_0(t)-\widehat{H}_0)
\widehat{R}_0(0)\nonumber\\
& &=\widehat{R}_0(t)(\tilde{a}(t)
\pd_x^2+\tilde{b}(t)\pd_x\pd_k)\widehat{R}_0(0)
\nonumber\\
& &\equiv \widehat{R}_0(t)B(t)
\eea
where $\tilde{a}(t)$ and $\tilde{b}(t)$ are both $C^{\infty}$ and 
$\mathcal{O}(t)$ as $t\ra 0$. 
It is proven in the Appendix, see Lemma \ref{relbound}, that
$\pd_x^2\widehat{R}_0(0)$ and 
$\pd_x\pd_k\widehat{R}_0(0)$ are bounded. Therefore the operator
$B(t)$ is bounded, $C^{\infty}$ and $B(t)\ra 0$ in norm as $t\ra 0$. 


With the properties of $B(t)$ listed above, we deduce that in a 
neighbourhood of $t=0$ 
\be
   \widehat{R}_0(t)=\widehat{R}_0(0)(\I -B(t))^{-1} 
\ee
which is $C^{\infty}$ in norm, since $B$ is, and we can conclude that
$H_0$ is $C^{\infty}(A)$.  
\\
To show that $(H_0+W)\in C^2(A)$ it is sufficient by
\cite{mourre}, \cite[Thm.~4.9]{cycon} and Lemma \ref{c1} to find some
$c>0$ such that
\be
(\ffi,[[W,iA],iA]\ffi)\leq c(\|H\ffi\|^2+\|\ffi\|^2)
\label{doublcomm}
\ee
for any $\ffi\in D(H)\cap D(A)$. Expanding the second
commutator in (\ref{doublcomm}) we write for any $\ffi\in
S(\R^2)$ 
\bea
&& (\ffi,[[W,iA],iA]\ffi) =
(\ffi,x(\pd_xW)\ffi)+(\ffi,x^2(\pd_x^2W)\ffi) \nonumber \\
&& +
i\,\mu\, [2(x(\pd_xW)\ffi,\pd_x\pd_y\ffi)-2(\pd_x\pd_y\ffi,x(\pd_xW)\ffi)]
\nonumber \\
&&
+i\mu\, [(\pd_x\pd_y\ffi,W\ffi)-(\ffi,W\pd_x\pd_y\ffi)]-\mu^2((\pd_x\pd_yW)\ffi,\pd_x\pd_y\ffi) \\ 
&&
-\mu^2[(\pd_x\pd_y\ffi,(\pd_x\pd_yW)\ffi)-(\pd_x,(\pd_y^2W)\pd_x\ffi)-(\pd_y\ffi,(\pd_x^2W)\pd_y\ffi)] \nonumber  
\eea
Now we can follow the proof of Lemma \ref{c1} and using the assumption
$(b)$ we get the following bound  
\bea
&& |(\ffi,[[W,iA],iA]\ffi)| \leq \|\ffi\|^2\|x^2\pd_x^2\,
W\|_{\infty}+W'_0\, \|\ffi\|(\|\ffi\|+4\|\pd_x\pd_y\ffi\|) \nonumber \\
&& +2\, \mu\, W_0\|\ffi\|\, \|\pd_x\pd_y\ffi\|
+\mu^2\|\pd_y^2W\|_{\infty}\|\pd_x\ffi\|^2 
\nonumber \\  
&&  +\mu^2\|\pd_x^2W\|_{\infty}\|\pd_y\ffi\|^2+
+2\, \mu^2\|\pd_x\pd_yW\|_{\infty}\|\pd_x\pd_y\ffi\|\,
\|\ffi\| \nonumber \\
&& \leq {\rm const}\, (\|H\ffi\|^2+\|\ffi\|^2)
\eea 
where the last inequality is justified by Lemma \ref{relbound}. \quad \QED

\vspace{0.3cm}

In order to prove the Mourre estimate (\ref{mest}) we will proceed in
two steps. First, we find a positive lower
bound on the contribution to the commutator coming from
$H_0$. Secondly, we control the contribution from $W$ so that
we preserve the sought positivity of $[H_0+W,iA]$. The former is done in

\vspace{2mm}

\begin{lem} \label{lowerb}
Let $\alpha>\delta>0$ and define
\be
I(\alpha,\delta):= \bigcup_{n\in{\bf N}_0}
[(2n+1)\alpha-\delta,(2n+1)\alpha+\delta]
\ee
Then for any $E\notin I(\alpha,\delta)$ there exists an
open interval $\Delta\ni E$ such that
$$
E_{\Delta}(H)[H_0,iA]E_{\Delta}(H) \geq \delta\,
E_{\Delta}(H)
$$
holds for $W_0$ small enough.
\end{lem}

{\it Proof:}
We define an operator
\be
H_L(\alpha) = \left (-i\pd_x\frac{B}{\alpha}+y\alpha\right
)^2- \pd_y^2
\ee
which is unitarily equivalent to the Landau Hamiltonian with the
magnetic field of a strength $\alpha$, so that
$\sigma(H_L(\alpha))=\{(2n+1)\alpha\}_{n\in{\bf N}_0}$. It follows that
\be
[H_0,iA]=-2\beta\, \pd_x^2= 2(H_0-H_L(\alpha))
\ee
Now, fix $\lambda\notin I(\alpha,\delta)$ and let us denote by
$n_0(\lambda)$ the largest 
natural number for which $\alpha(2n_0(\lambda)+1)\leq\lambda$.
The spectral family of $H_0$ is thus given by
\be
E_0(\lambda)=\sum_{n\leq n_0(\lambda)}P_n\,
\chi_t([0,\lambda-\alpha(2n+1)))
\ee
where $P_n$ is the projection on the $n^{th}$ Landau level of
$H_L(\alpha)$ and $\chi_t$ is the spectral projection of $-\beta\, \pd_x^2$.

To continue consider  an open interval
$\tilde{\Delta}=(E-\eps,E+\eps)$ with
$\eps$ such that
$\tilde{\Delta}\not\subset I(\alpha,\delta)$.
For the spectral projection of $H_0$ on the interval
$\tilde{\Delta}$ we then get
\bea
E_{\tilde{\Delta}}(H_0) & = & E_0(E+\eps)-E_0(E-\eps) \nonumber \\
& = & \sum_{n\leq
  n_0(E)}P_n\, \chi_t([E-(2n+1)\alpha-\eps,E-(2n+1)\alpha+\eps))
\eea
and this gives us the lower bound on
$E_{\tilde{\Delta}}(H_0)[H_0,iA]E_{\tilde{\Delta}}(H_0)$ in the form
\bea 
&& E_{\tilde{\Delta}}(H_0)[H_0,iA]E_{\tilde{\Delta}}(H_0) = 
E_{\tilde{\Delta}}(H_0)(-2\beta\, \pd_x^2)E_{\tilde{\Delta}}(H_0) \nonumber \\ 
&& = \sum_{n\leq
  n_0(E)}P_n\,
\chi_t([E-(2n+1)\alpha-\eps,E-(2n+1)\alpha+\eps)))(-2\beta\, \pd_x^2)
\nonumber \\
&& P_n\,
\chi_t([E-(2n+1)\alpha-\eps,E-(2n+1)\alpha+\eps))) \geq  
E_{\tilde{\Delta}}(H_0)\, 2\, \delta 
\eea
Applying the argument of \cite{fro} this result can be extended to
$H$. For $I(\alpha,\delta)\not\supset\Delta\ni E$ we decompose $E_{\Delta}(H)$ as
$$
E_{\Delta}(H)=E_{\tilde{\Delta}}(H_0)E_{\Delta}(H)
+(1-E_{\tilde{\Delta}}(H_0))E_{\Delta}(H)
$$
and since $E_{\tilde{\Delta}}(H_0)$ commutes with $[H_0,iA]$ we get 
\bea 
&& E_{\Delta}(H)\left ([H_0,iA]-2\,
\delta\right )E_{\Delta}(H) =  \nonumber \\
&& = E_{\Delta}(H)E_{\tilde{\Delta}}(H_0)([H_0,iA]-2\,
\delta)E_{\tilde{\Delta}}(H_0)E_{\Delta}(H) \nonumber \\ 
&& + E_{\Delta}(H)([H_0,iA]-2\,
\delta)(1-E_{\tilde{\Delta}}(H_0))E_{\Delta}(H) 
\eea
From this one easily obtains the following inequality 
\bea
&& E_{\Delta}(H)\left ([H_0,iA]-2\,
\delta\right )E_{\Delta}(H) \nonumber \\
&& \geq
E_{\Delta}(H)E_{\tilde{\Delta}}(H_0)([H_0,iA]-2\,
\delta)E_{\tilde{\Delta}}(H_0)E_{\Delta}(H) \nonumber \\ 
&& - \|([H_0,iA]-2\,
\delta)(1-E_{\tilde{\Delta}}(H_0))E_{\Delta}(H)\| 
\eea
where the first term on the r.h.s. is non-negative. 
From Lemma \ref{relbound} we know that 
\be
\|\beta\, \pd_x^2\, H_0^{-1}\|\leq \beta\, C(\omega,B) =c\,
\frac{1+\alpha^2}{\alpha^2} 
\ee
where $c$ is a numerical constant.
We can thus follow \cite{fro} and claim that the second term is
bounded from above by
\bea
&& 2\, \beta\, C(\omega,B)\|H_0(1-E_{\tilde{\Delta}}(H_0))(H_0-E)^{-1}\|\,
\|(H_0-E)E_{\Delta}(H)\| \nonumber \\
&& +2\, \delta\, \|H_0^{-1}\|\, \|H_0(1-E_{\tilde{\Delta}}(H_0))(H_0-E)^{-1}\|\,
\|(H_0-E)E_{\Delta}(H)\| \nonumber \\
&&  \leq 2(\delta\, \alpha^{-1}+ \beta\, C(\omega,B))(1+E\,\eps^{-1})(|\Delta|+W_0)
\eea 
so that for 
\be
(|\Delta|+W_0)< \frac{\delta}{2(\delta\, \alpha^{-1}+ \beta\,
  C(\omega,B))(1+E\eps^{-1})} 
\ee
is
$$
E_{\Delta}(H)([H_0,iA]-2\,
\delta)E_{\Delta}(H)\geq -\delta
$$
and hence
\be
E_{\Delta}(H)[H_0,iA]E_{\Delta}(H)\geq
\delta E_{\Delta}(H)
\ee
what we set out to prove. \quad \QED

\vspace{0.7cm}

Armed with these Lemmas we are in position to prove the Mourre
estimate for $H$. 

\vspace{2mm}

\begin{lem} \label{ac}
Let $E\notin I(\alpha,\delta+\eps)$. Assume moreover that
\bea
&& (I)\quad W_0 <  \frac{\delta}{2(\delta\, \alpha^{-1}+ \beta\,
  C(\omega,B))(1+E\eps^{-1})} \nonumber \\
& {\rm and} & \label{assum}  \\
&& (II)\quad W'_0+B\, \alpha^{-2}\, \sqrt{c\, C(\omega,B)}\, W_0\, (E+W_0) < \delta/2
\nonumber
\eea
Then there is an open interval $\Delta\ni E$ such
that
\be
E_{\Delta}(H)[H,iA]E_{\Delta}(H)\geq \delta/2\, E_{\Delta}(H)
\ee
\end{lem}

\begin{figure}[t]
\setlength{\unitlength}{1mm}
\begin{picture}(50,60)
\linethickness{0.5mm}
\put(30,55){\line(1,0){70}}
\put(30,0){\line(1,0){70}}
\linethickness{0.2mm}
\put(102,54){$(2n+3)\, \alpha$}
\put(102,-1){$(2n+1)\, \alpha$}
\put(30,47){\line(1,0){70}}
\put(30,8){\line(1,0){70}}
\linethickness{0.5mm}
\put(30,40){\line(1,0){70}}
\put(30,15){\line(1,0){70}}
\linethickness{0.2mm}
\put(35,55){\vector(0,-1){8}}
\put(37,50){$\delta$} 
\put(35,0){\vector(0,1){8}}
\put(37,3){$\delta$} 
\put(95,47){\vector(0,-1){7}}
\put(97,42){$\eps$} 
\put(95,8){\vector(0,1){7}}
\put(97,11){$\eps$}
\put(64,27){\line(1,0){2}}
\put(67,25.5){$E$} 
\put(65,40){\line(0,-1){25}}
\put(65,20){\line(-1,0){17}}
\put(65,30){\line(-1,0){17}}
\put(80,35){\line(-1,0){15}}
\put(80,23){\line(-1,0){22}}
\put(50,25){\vector(0,-1){5}}
\put(50,25){\vector(0,1){5}}
\put(43,24){$\Delta_1$}
\put(78,29){\vector(0,-1){6}}
\put(78,29){\vector(0,1){6}}
\put(80,28){$\Delta_2$}
\linethickness{0.4mm}
\put(60,26.5){\vector(0,-1){3.5}}
\put(60,26.5){\vector(0,1){3.5}}
\linethickness{0.2mm}
\put(55,25){$\Delta$} 
\end{picture}
\caption{Energy intervals for the Mourre estimate \label{fig1}}
\end{figure}
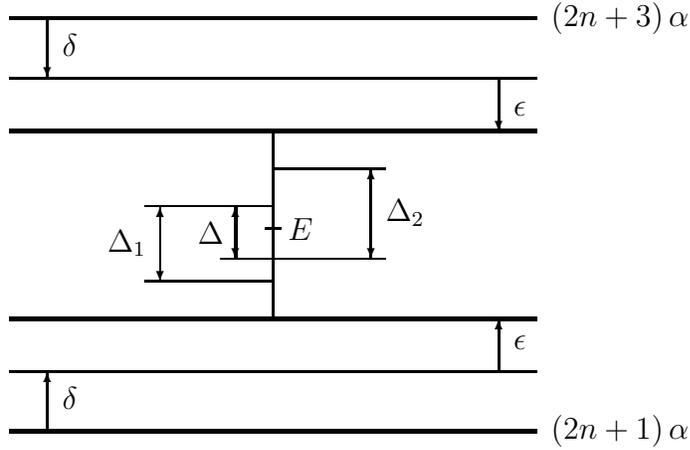

{\it Proof:} Consider again some open interval $\Delta_1\ni E$, see
Fig. \ref{fig1}, and a state
$\psi=E_{\Delta_1}(H)\psi$. We mimick the argument used in the
proof of Lemma \ref{c1} and keeping in mind that
$\|(H-E)\psi\|\leq |\Delta_1|\, \|\psi\|$ we get
\bea 
|(\psi,[W,iA]\psi )| &\leq & W'_0\, \|\psi\|^2 +
2B\alpha^{-2}W_0\|\pd_x\pd_y\psi\|\, \|\psi\| \\ 
&\leq & W'_0\, \|\psi\|^2+B\alpha^{-2}\sqrt{c\, C(\omega,B)}\,
W_0(E+W_0+|\Delta_1|)\, \|\psi\|^2 \nonumber
\eea 
where we have used the fact that $2\|\pd_x\pd_yH_0^{-1}\|\leq
\sqrt{c\, C(\omega,B)}$, see Lemma \ref{relbound}.

By letting $|\Delta_1|\ra 0$ we get from (\ref{assum}) the upper
bound on the contribution from $W(x,y)$:
\be
|(\psi,[W,iA]\psi)|< \delta/2\, \|\psi\|^2
\ee


On the other hand by Lemma \ref{lowerb} for $W_0$ sufficiently small
there is $\Delta_2\ni E$ such that
\be
(\psi,[H_0,iA]\psi)\geq
\delta\, \|\psi\|^2
\ee
for $\psi=E_{\Delta_2}(H)\psi$.

To complete the proof it sufficies to take $\Delta=\Delta_1\cap\Delta_2$.
\quad \QED

\vspace{0.7cm}

Note that once the condition (\ref{assum}) holds for
some $\tilde{E}$, it holds also for all $E\leq \tilde{E}$. This leads us to the
following definition:
\be
\Delta(E,\alpha,\delta+\eps):= \{\lambda |\, \lambda\leq E,\, \lambda\notin
I(\alpha,\delta+\eps) \}
\ee
Now we are ready to state our main result.

\vspace{2mm}
 
\begin{thm} \label{main}
Assume $W_0 \!=\! \|W\|_{\infty}<\alpha,\, 
W'_0=\|x\, \pd_xW\|_{\infty}<\infty $ and that the assupmtions of
Lemma $(\ref{ac})$ are satisfied
for some $\eps$ and $E\notin I(\alpha,\delta+\eps)$. Then
\begin{itemize}
\item[$(1)$] $H$ has no eigenvalues in the interval
  $\Delta(E,\alpha,\delta+\eps)$,
\item[$(2)$] if in addition $W\in C^2(\R^2)$ and  
$$
 \|\pd_x^2W\|_{\infty}<\infty,\,
  \|\pd_y^2W\|_{\infty}<\infty,\, \|\pd_x\pd_yW\|_{\infty}<\infty,\,
  \|x^2\pd_x^2W\|_{\infty}<\infty,
$$
then the spectrum of $H$ in the interval $\Delta(E,\alpha,\delta+\eps)$
is purely absolutely continuous.
\end{itemize}
\end{thm}

\vspace{0.3cm}

{\it Proof:} Application of the Virial
respectively Mourre Theorem and Lemmas \ref{c1}, \ref{c1+a}
and \ref{ac}.

\vspace{0.3cm}

\noindent {\bf Remark:} Theorem \ref{main} does not exclude the possibility
that the spectrum of $H$ is empty in the considered interval. However,
it follows from the standard perturbative argument that since the
spectrum of $H_0=H-W$ includes whole the interval 
$[\alpha,\infty)$ this cannot happen for $W_0$ small enough.

\vspace{0.3cm}

Let us now consider the following scaling:
$$
E=E_0\, \alpha,\quad \delta=\delta_0\, \alpha, \quad \eps = \eps_0\,
\alpha
$$ 
where $E_0,\, \delta_0,\, \eps_0$ are fixed. From $(I)$ we then get 
\be
W_0 < \frac{\delta_0\, \eps_0\,
 \alpha}{2(\delta_0+c\frac{1+\alpha^2}{\alpha^2})(\eps_0+E_0)}\ra\infty, \quad {\rm as} \quad \omega\ra\infty  
\ee
and similarly from $(II)$
\be
W'_0 < \alpha\,
\delta_0/2-cB\alpha^{-2}\sqrt{\frac{1+\alpha^2}{\alpha^2}}\, W_0(E_0\,
\alpha+W_0)\ra\infty, \quad {\rm as} \quad \omega\ra\infty
\ee
In other words, for $\omega$ sufficiently large there is some
interval in between the modified Landau levels, in which
the transport survives whenever $W_0,\, W'_0 <\infty$. We thus have

\vspace{2mm}

\begin{cor} Let $E_0,\, \delta_0,\, \eps_0$ be fixed and assume that
both $W_0$ and $W'_0$ are finite. Then the statements
of Theorem $\ref{main}$ hold in the interval $\Delta(\alpha\, E_0,
\alpha(\delta_0+\eps_0))$ provided $\omega$ is large enough.
\end{cor}

On the other hand, in the high energy limit the behaviour of the bound
$(I)$ is as $E^{-1}$. Accordingly, Theorem \ref{main} proves the
absence of eigenvalues respectively absolute continuity only in a finite
number of intervals. In this sence our result is comparable with
those of \cite{fro,biev}, where the upper bound on the size of perturbation
is also $\mathcal{O}(E^{-1})$ as $E\ra\infty$.   For comparison we note
that the same bound on $\|W\|_{\infty}$ obtained in
\cite{macr} is decreasing with energy as $E^{-4}$.


\subsection{The positivity of $[H_0,iA]$: more general approach} \label{gen}

As we have seen above, the condition $W'_0<\infty$ which doesn't allow us
to consider non-localized perturbations, e.g. random, comes from the
fact that our conjugate operator includes the dilation generator $x\,
p_x$. Let us now show that, for $A$ being a quadratic function of
$(x,y,p_x,p_y)$, the presence of this term is necessary if one
requires $[H_0,iA]$ to be definitly positive.\par
We take $A$ in the form
\bea A &=& \sum_{j,k} \alpha_{j,k}\pd_{x_j}\pd_{x_k}+
i\sum_{j,k}\beta_{j,k}(x_k\pd_{x_j}+\pd_{x_j}x_k) \nonumber \\ &+&
\sum_{j,k} \gamma_{j,k}x_j x_k + i \sum_j \delta_j\, \pd_{x_j}+
\sum_j\eps_j x_j \eea
where $j,k=1,2$. Assume that the ``bad'' term is absent, i.e.
$\beta_{1,1}=0$. The straightforward computation then gives
 \bea
[H_0,iA] & = & 4B\alpha_{1,2}\,
p_1^2+2(B\alpha_{2,2}-\beta_{1,2}-\beta_{2,1})\,
p_1p_2-4\beta_{2,2}\, p_2^2 +4\gamma_{1,2}\, x_1p_2 \nonumber \\ &
+ & (2\gamma_{1,1}+B\beta_{2,1})(x_1p_1+p_1x_1)
+4(\alpha^2\alpha_{1,2}+\gamma_{1,2}-B\beta_{2,2})\, x_2p_1
\nonumber \\ & + &
2(2\alpha^2\alpha_{2,2}+2\gamma_{2,2}-B\beta_{2,1})(x_2p_2+p_2x_2)
\nonumber \\ & + & 4(\alpha^2\beta_{2,1}+B\gamma_{1,1})\,
x_1x_2+4(\alpha^2\beta_{2,2}+B\gamma_{1,2})\, x_2^2
+i(\eps_1p_1+\eps_2p_2) \nonumber \\ & + & 2\delta_2\alpha^2\,
x_2- 2i(\gamma_{1,1}+\gamma_{2,2}+\alpha^2\alpha_{2,2}) \eea
First of all notice that since $H_0$ is purely quadratic, the
linear terms of $A$ produce again only linear terms in $[H_0,iA]$
and we can thus leave them out without loss of generality. The
cenral point is that, due to the translation invariance in $x$,
the term proportional to $x_1^2$ is missing in $[H_0,iA]$. This
means that if we want $[H_0,iA]$ to be definitely positive, we
have to make the terms with $x_1$ vanish:
\be
\gamma_{1,2}=0,\quad 2\gamma_{1,1}+B\beta_{2,1}=0,\quad
\alpha^2\beta_{2,1}+B\gamma_{1,1}=0
\ee
But now $x_2^2$ and $p_2^2$ have necessarily opposite signs, so that
we need also $\beta_{2,2}$ to be zero, which implies that $x_2^2$ is
absent as well. Following the argument given above for $x_1^2$ we get
\be
\alpha_{1,2}=0,\quad 2\alpha^2\alpha_{2,2}+2\gamma_{2,2}-B\beta_{2,1}=0
\ee
and we are left with
$$
2(B\alpha_{2,2}-\beta_{1,2}-\beta_{2,1})\,
p_1p_2
$$
which cannot be definite positive.

\section*{Appendix}

{\it Proof of Lemma $\ref{relbound}$:} Application of a partial Fourier
transform in $x$ shows that $H_0$ is unitarily equivalent to
\be
\hat{H_0} = -\pd_v^2+u^2+2Buv+\alpha^2\, v^2 = P^2 + V(u,v)
\label{def}
\ee
where $P := -i\pd_v$. We now mimick the argument used in
\cite[Ex.~7.2.4]{beh}. First of all note that since
$$
u^2+2Buv+\alpha^2v^2 = (u+Bv)^2+\omega^2v^2
$$
we can write
$$
V(u,v)=(V^{1/2}(u,v))^2
$$
For $\psi\in S(\R^2)$
\bea
&& \|(P^2+V)\psi\|^2=(\psi,(P^4+V^2+P^2V+VP^2)\psi) \nonumber \\
&& = (\psi,(P^4+V^2+2PVP+[P,[P,V]])\psi)
\eea
Furthermore, we compute
$$
[P,[P,V]]=[P,-i\pd_v\, V]=-\pd_v^2 V= -2\alpha^2
$$    
Then
$$
\|(P^2+V)\psi\|^2=\|P^2\psi\|^2+\|V\psi\|^2+2\|V^{1/2}P\psi\|^2-2\alpha^2\|\psi\|^2
$$
so that
$$
\|P^2\psi\|^2+\|V\psi\|^2 \leq 2\alpha^2\|\psi\|^2+\|(P^2+V)\psi\|^2
$$
Since both $P^2,\, V$ are closed we can follow the argument given in
\cite[Ex.~7.2.4]{beh} and claim that
\be
D(P^2+V) = D(P^2)\cap D(V)
\ee
Taking $\hat{R_0}(\lambda)=(\hat{H_0}+\lambda)^{-1}$ for some $\lambda>0$ 
it then follows from closed graph Theorem that both 
$$
P^2 \hat{R_0}(\lambda),\quad V \hat{R_0}(\lambda)
$$
are bounded. More precisely, one can show that for any $\psi\in S(\R^2)$ 
\be
\|P^2 \hat{R_0}(\lambda)\psi\|\leq \sqrt{6}\, \|\psi\|, \quad \|V
\hat{R_0}(\lambda)\psi\|\leq \sqrt{6}\, \|\psi\|
\ee
which proves $(i)$
To continue we note that $V(u,v)$ can be diagonalized by an orthogonal
transform $T$ so that
\be
V(u,v) = \lambda_{+}\hat{u}^2+\lambda_{-}\hat{v}^2
\ee
where $(\hat{u},\hat{v})=T(u,v)$ and
$$
\lambda_{\pm}=\frac{1+\alpha^2\pm\sqrt{(1+\alpha^2)^2-4\omega^2}}{2}
$$
Therefore we have
\bea
V(u,v)& \geq & \lambda_{-}(u^2+v^2)=\frac{1}{2}\,
\frac{(1+\alpha^2)^2-(1+\alpha^2)^2+4\omega^2}{1+\alpha^2+\sqrt{(1+\alpha^2)^2-4\omega^2}}\, (u^2+v^2)
  \nonumber \\
& \geq & \frac{\omega^2}{1+\alpha^2}\, (u^2+v^2)
\eea
From (\ref{def}) we know that there exists a unitary operator $U$ such that
$$
\hat{H_0} = U H_0 U^{-1}
$$
Now taking $\ffi=U\psi$ we get
\be
\|\pd_x^2\psi\|=\|u^2\ffi\|\leq
\frac{1+\alpha^2}{\omega^2}\|V\ffi\|, \, \|y^2\psi\|=\|v^2\ffi\|\leq
\frac{1+\alpha^2}{\omega^2}\|V\ffi\|
\ee
and
\be
\|y\pd_x\psi\|=\|uv\ffi\|\leq \frac{1}{2}\,
\frac{1+\alpha^2}{\omega^2}\|V\ffi\|
\ee
which gives us $(ii)$. Finally, 
\be
\|\pd_x\pd_y\psi\|^2 = (u\, P\ffi,u\, P\ffi)\leq \|P^2\ffi\|\,
\|u^2\ffi\|\leq c^2\, \frac{1+\alpha^2}{\omega^2}\, \|\hat{H_0} \ffi\|^2 
\ee
\QED

\section*{Acknowledgement} 
Useful discussions with  J.-M.~Combes, N.~Macris, 
and Ph.~Martin
are gratefully acknowledged. A.~J. thanks the Doppler Institute, Czech
Technical University, where this work has
begun, for the hospitality. H.~K. would like to thank his hosts at
Institut Fourier in Grenoble for the warm hospitality extended to him. 
The research has been partially supported by the Grant Agency of the
Czech Academy of Sciences under the Contract A1048101 and by the program
Tempra from R\' egion Rh\^one-Alpes.

\end{document}